\documentclass{article}
\usepackage[utf8]{inputenc}
\usepackage{spconf,amsmath,graphicx}

\title{MUSIC PLAGIARISM DETECTION: PROBLEM FORMULATION AND A SEGMENT-BASED SOLUTION}
\name{Seonghyeon Go$^{\star}$, Yumin Kim$^{\star}$\thanks{$^{\star}$Equal Contribution}}

\address{MIPPIA Inc.}

\begin{document}
%
\maketitle
\begin{abstract}

Recently, the problem of music plagiarism has emerged as an even more pressing social issue. As music information retrieval research advances, there is a growing effort to address issues related to music plagiarism. However, many studies, including our previous work, have conducted research without clearly defining what the music plagiarism detection task actually involves. This lack of a clear definition has slowed research progress and made it hard to apply results to real-world scenarios. To fix this situation, we defined how Music Plagiarism Detection is different from other MIR tasks and explained what problems need to be solved. We introduce the Similar Music Pair dataset to support this newly defined task. In addition, we propose a method based on segment transcription as one way to solve the task. Our demo and dataset are available at https://github.com/Mippia/ICASSP2026-MPD.


\end{abstract}
\section{Introduction}
\label{sec:intro}
Music plagiarism is becoming more of a social issue~\cite{yuan2023perceptual}. Recently, many studies have been conducted to prevent music plagiarism, but the research findings often show some inconsistencies. In most cases, research was conducted without clarifying what a music plagiarism task is. This phenomenon has even been revealed in the previous work of this paper~\cite{go2025real}. Also, the datasets are often far from the actual plagiarism. 

Various studies have recently investigated on the Music Plagiarism Detection task. A Siamese Network approach that modifies melodies has been proposed~\cite{park2022music}, and there are also cases where similar data were constructed across both MIDI and audio~\cite{lu2025melodysim}. In addition, some studies applied NLP-based methods with tokenization~\cite{malandrino2022adaptive}, while others adopted audio fingerprinting-based approaches~\cite{borkar2021music}.

However, these methodologies share a common limitation in that they rely on artificially constructed datasets. Models trained on such data are likely to overfit to computable solutions, which limits their applicability to real-world scenarios. In contrast, a study based on an actual plagiarism case has also been reported~\cite{liu2023fine}. Nevertheless, this work was restricted only to MIDI data. Furthermore, although all of the above studies were conducted under the task of ``Music Plagiarism Detection'', the datasets and evaluation metrics used are inconsistent, making direct comparison difficult.

To solve these problems, we first define what the Music Plagiarism Detection task is and explain how it differs from the existing music similarity tasks, such as Audio Fingerprinting and Cover Song Identification. We propose a method that uses Segment Transcription to divide existing music into small indexed segments, creating a searchable library of music database. Our approach uses a multimodal siamese network that leverages both audio and pianoroll representations to retrieve potentially plagiarized data from the library for a given query music. Our contributions are summarized as:

\begin{itemize}
\item We provide a clear definition of the Music Plagiarism Detection task and distinguish it from other Music Information Retrieval tasks.
\item We employ a segment transcription methodology that converts raw audio into structured musical representations optimized for plagiarism detection.
\item We evaluated various plagiarism detection models based on similarity networks for effective music representation learning and similarity analysis.

\end{itemize}
\begin{figure*}[t]
\centering
\includegraphics[width=\textwidth]{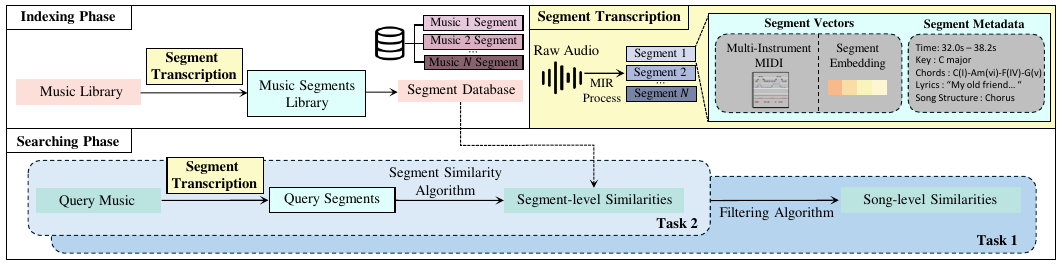}
\caption{Overall architecture of the proposed music plagiarism detection system. 
The pipeline processes input audio through segment transcription, extracts musical features 
for each instrument and structural component, and applies both algorithmic and deep learning-based 
similarity analysis to detect plagiarism cases.}
\label{fig:system_overview}
\end{figure*}

\section{Music Plagiarism Detection}
\label{sec:format}

\subsection{Task Description}
Music plagiarism is an act of infringing on rights by copying the music work of others. However, recent disputes over music plagiarism do not arise only when it is confirmed that the plagiarist is intentionally ingrained. It may appear to be a similar sound source by coincidence, and the problem is even occurring due to the development of generative AI. For this reason, the Music Plagiarism Detection task is important for all groups in the music ecosystem, including record labels, artists, and even those accused of plagiarism.

Music Plagiarism Detection is somewhat associated with other music similarity tasks in the field of Music Information Retrieval (MIR). Cover Song Identification (CSI) is the task of finding a cover version of the original song and determining musical similarity in the entire music. Melody MIDI-based methodology~\cite{marolt2006mid} or methodology that interprets and compares sound sources as sequences~\cite{serra2009cross} have been traditionally used. Recently, CNN-based methods~\cite{ du2023bytecover3} and models using a conformer structure~\cite{liu2023coverhunter} are achieving SOTA performance. And another similar task, Audio Fingerprinting~\cite{chang2021neural} is a technology that extracts unique features from short audio samples to identify entire songs, and is mainly used for song identification in music streaming services or broadcasts.

However, Music Plagiarism Detection differs from these existing music similarity tasks in quite a few ways.
First, it is characterized by partial similarity. Some music retrieval methods, like CSI, work by judging similarity only at the whole music-level, while music plagiarism may occur only in a specific section of the music. It is difficult to detect plagiarism by comparing the entire music if only a short part is copied. As a result, plagiarism detection requires a precise comparative analysis for each section of music.

Second, it is characterized by selective music elements. Some musical similarity methods, like audio fingerprinting, focus on acoustic signal characteristics. But plagiarism can only appear in specific elements of the signal. For example, if the mixing, mastering, and instruments are completely different but only the topline melody is the same, it is difficult to detect plagiarism by signal similarity computation.

Due to these differences, the plagiarism detection task requires a different approach from existing music similarity methods. We define the Music Plagiarism Detection task as follows. 
When a full-length music A is given as a Query:
\begin{itemize}
  \item \textbf{Task 1)} The model should be able to select $A'$, which plagiarizes $A$ from the large database.
  \item \textbf{Task 2)} The model should be able to clearly specify which part of $A$ is similar to which part of $A'$.
  \item \textbf{Task 3)} If possible, the model should explain why the part is plagiarized (melody, vocal, or chord).
\end{itemize}


\section{Method}





Tasks 1, 2, and 3 can be accomplished in any order as the system is capable of performing each task effectively. Our proposed Music Plagiarism Detection system based on the standard query–search methodology ~\cite{gaillard2017large}, which integrates Music Segment Transcription with similarity analysis. Figure~\ref{fig:system_overview} shows the overall architecture of our approach, which consists of three main stages: (1) Music Segment Transcription that converts raw audio into structured musical representations. (2) Segment-level similarity analysis is employed to perform Task 2. (3) Filtering music with segment-level similarities to perform Task 1. Task 3 can be derived depending on each similarity algorithm.


\subsection{Music Segment Transcription}
Using the Music Segment Transcription system from our previous study~\cite{go2025real}, we extract musical segments from audio. Each segment represents a unit containing essential musical information. Specifically, we used Demucs~\cite{rouard2023hybrid} for source separation, the all-in-one model~\cite{kim2023all} for structural analysis, AST~\cite{wang2021preparation} for vocal transcription, SheetSage~\cite{donahue2022melody} for melody transcription, and Harmony Transformer~\cite{chen2019harmony} for chord transcription. Depending on how much information the segment should contain, models can be added or removed. This system converts raw audio into structured music representations.



\begin{table*}[t]
\centering

\begin{tabular}{|l|l|l|l|l|l|l|}
\hline
\textbf{Original Title} & \textbf{Comparison Title} & \textbf{Relation} & \textbf{Original Time} & \textbf{Comparison Time} & \textbf{Pair \#} & \textbf{Acoustic index}\\
\hline

Shiki no uta & Bul-ggot & Plagiarism & [73, 82, 134, ...] & [85, 93, 136, ...] & 25 & 52\\
\hline
I Will Survive & I Will Survive(Jinju) & Remake & [8, 24, 41, ...] & [18, 34, 66, ...] & 62 & 135\\
\hline
I Will Survive & I Will Survice(Jinju) & Remake & [16, 33, 50, ...] & [26, 42, 74, ...] & 62 & 136\\
\hline
\end{tabular}
\caption{Sample entries from the Similar Music Pair dataset}
\label{tab:dataset_sample}
\end{table*}

\subsection{Music Segment Similarity}
It is essential to accurately measure the similarity between segments for plagiarism detection. Segment similarity measurement should be performed at various levels, from analysis of individual music elements to recognition of overall musical patterns by the model. In this paper, two different approaches are used to compare segments.

\noindent\textbf{Music domain-based similarity measurement.} The first approach is a similarity measurement based on musical domain knowledge proposed in the previous study~\cite{go2025real}. We compute  pianoroll similarity, onset rhythm similarity, and chord similarity with segment information. Each element is calculated independently to derive the final similarity score through weighted combinations. Through this analysis, it is possible to perform \textbf{Task 3} because it is possible to clearly explain which elements of music are similar. However, regardless of performance, the algorithmic method is difficult to prove its validity, and it is quite outdated in recent research directions.

\noindent\textbf{Deep Learning-Based Similarity Measurement.} The second approach employs a deep learning model that automatically learns plagiarism patterns from audio data. We individually fine-tune MERT~\cite{li2023mert} and trained pianoroll CNN~\cite{park2022music} models with Siamese network architecture, and tried a dual-encoder cross-fusion method that combines two pre-training models with bidirectional cross-attention.
 MERT's acoustic feature-based approach may face the same selective element challenges as Audio Fingerprinting. That is, plagiarism in specific musical components can be masked by different acoustic features. Music Segment Transcription and CNN-based methods are relatively robust to these limitations. However, we include the MERT experiment to provide a comprehensive comparison. 

\section{Experiment}

\subsection{Dataset}
Table~\ref{tab:dataset_sample} shows a sample of SMP datasets. This is a comprehensive music piece pair dataset for plagiarism detection evaluation, which was constructed for the experiments. It consists of 72 pairs of original songs and comparative songs, each containing time annotations that mark where similar segments begin. Since different sound patterns can appear even within the same song pair, we constructed these parts separately and expressed them as an acoustic index. The dataset consists of actual plagiarism and remake cases, and contains accurate time information of similar parts for each case. This dataset encompasses various music genres and similarity types, enabling a comprehensive evaluation of diverse scenarios arising from real music copyright disputes.

In addition, we used the Covers80~\cite{ellis2007covers80} dataset. It consists of 80 cover song groups, with each group including multiple versions of the same original song performed by different artists. This dataset was used for the establishment of the unseen segment database and for performance comparison with music-level experiments conducted in existing CSI works.

\subsection{Experiment Settings}
Our experiment was conducted using a single NVIDIA GeForce RTX 5090 GPU. Learning rate 1e-4, batch size 32, and 100 epochs were used as basic settings for model learning, and the Adam optimizer was used. In the case of deep learning-based similarity experiments, we use 5-fold training due to the relatively small size of the dataset. All segments used in the experiment have a 4-bar length.

\subsection{Evaluation Metrics}
For segment-level experiments related to \textbf{Task 2}, we employ a temporal-precise recall metric. Rec. 1s@k measures recall at top-k when retrieved segment timestamps are within 1 second of ground truth compare times in the same acoustic index.

The validation set in fold (14 comparison music pieces) and music pieces in Covers80 (160 music pieces) are given as the index library. Each metric is computed across three configurations: SMP Timestamps (14 music pieces, avg 149 indices) using only annotated timestamp segments from the SMP dataset, SMP (14 music pieces, avg 1,525 indices) using all extractable segments from the SMP dataset songs, whose start points are all downbeats extracted by the beat-tracking model, and Full Indices (174 music pieces, avg 24,843 indices), incorporating all extractable segments from the complete dataset, including Covers80.

To evaluate the system's capability for detecting plagiarism without prior knowledge of query segments, we evaluate music-level performance related to \textbf{Task 1}. The final decision is made through a filtering algorithm that applies weighted majority voting among the top-20 segment-level matches. Weights decrease linearly from 20 for rank 1 to 1 for rank 20 to capture cases where only partial musical content is similar. we used mean average precision (mAP) and mean rank of the first correct result (MR1) metrics, widely used in CSI tasks.

\subsection{Experimental Results}

\subsubsection{Segment-Level Performance}

Table~\ref{tab:segment_performance} presents segment-level performance results across different evaluation settings. Segment-level plagiarism detection presents considerable challenges due to the difficulty of accurately pinpointing the exact temporal boundaries of plagiarized content. Models may retrieve segments from ground-truth music but fail to precisely locate the specific sections, making the music plagiarism detection task difficult.

MERT~\cite{li2023mert} shows relatively high performance across training models, benefiting from its music-specific pre-trained representations. Multimodal models underperformed despite high expectations, possibly due to insufficient training data.

The music domain-based approach (Music) shows distinctive characteristics, maintaining relatively stable performance patterns across different scales.

Performance degradation across increasing dataset scales demonstrates the scalability challenges in large-scale music similarity search. Despite the relatively small number of audio tracks in this experiment, the available segment count was considerably large. Future work could focus on either refining segment selection strategies or developing more robust models that can maintain performance at scale.
\begin{table}[t]
\centering
\resizebox{\columnwidth}{!}{%
\begin{tabular}{l|ccc|ccc|ccc}
\hline
& \multicolumn{3}{c|}{\textbf{SMP Timestamps}} & \multicolumn{3}{c|}{\textbf{SMP}} & \multicolumn{3}{c}{\textbf{Full Indices}} \\
\cline{2-10}
 & \multicolumn{3}{c|}{Rec.1s} & \multicolumn{3}{c|}{Rec.1s} & \multicolumn{3}{c}{Rec.1s} \\
\textbf{Method}  & @1 & @5 & @10 & @1 & @5 & @10 & @1 & @5 & @10 \\
\hline
CNN & 13.4 & 32.1 & 45.9 & 2.9 & 8.9 & 14.3 & 1.5 & 3.1 & 4.2 \\
MERT & \textbf{25.6} & \textbf{43.0} & \textbf{51.0} & 4.3 & 11.1 & 14.3 & 2.2 & 7.0 & 7.9 \\
MM & 20.2 & 33.7 & 41.5 & 3.2 & 8.9 & 11.3 & 2.1 & 5.2 & 5.7 \\
Music & 7.8 & 23.4 & 29.6 & \textbf{5.8} & \textbf{19.6} & \textbf{26.1} & \textbf{4.0} & \textbf{12.3} & \textbf{15.4} \\
\hline
\end{tabular}%
}

\caption{Segment-Level Music Plagiarism Detection Results, MM stands for Multi-Modal, Music stands for music domain-based method}
\label{tab:segment_performance}
\end{table}

\subsubsection{Music-level Performance}

Table \ref{tab:music_performance} presents the music-level performance results on the Covers80 dataset. Our approach shows lower performance compared to state-of-the-art CSI models such as Bytecover3~\cite{du2023bytecover3} and CoverHunter~\cite{liu2023coverhunter}. This performance difference reflects the fundamental distinction between cover song identification and music plagiarism detection tasks. CSI models are designed to identify broad musical similarities across entire songs, optimizing for overall music-level matching. In contrast, music plagiarism detection requires precise identification of specific temporal segments, demanding a different set of capabilities. While this does not suggest that our system is in an optimal state, we expect that our approach provides unique advantages in plagiarism detection applications.

\begin{table}[b]
\centering

\begin{tabular}{|l|c|c|}
\hline
\textbf{Method} & \textbf{MAP $\uparrow$ } & \textbf{MR1 $\downarrow$ } \\
\hline
Bytecover3~\cite{du2023bytecover3} & 0.927& 3.32\\
CoverHunter~\cite{liu2023coverhunter} & \textbf{0.933} & \textbf{3.20}\\
MERT & 0.15 & 33.3  \\
Multimodal & 0.2 & 28.8  \\
CNN & 0.11 & 46.6  \\
Music & 0.475 & 13.46 \\
\hline
\end{tabular}
\caption{Music-Level Performance in Covers80 dataset}
\label{tab:music_performance}
\end{table}

Our approach demonstrates a key strength in providing precise segment matching with exact temporal localization. Table~\ref{tab:segment_examples} shows examples of music-level detection drafts in the Covers80 dataset. The system can pinpoint specific musical segments while providing interpretable reasoning.

This represents a significant aspect that distinguishes our work from traditional CSI approaches. While CSI models show superior performance at broad similarity detection, they may not provide the exact timing and duration of copied material. Our system enables detailed forensic analysis by identifying specific musical segments with their corresponding reasoning. Additionally, the system maintains explainability even in incorrect cases, providing specific musical justifications that can be evaluated. 


\begin{table}[t]
\centering
\begin{tabular}{l|l}
\hline
\multicolumn{2}{c}{\textbf{Correct Example 1}} \\
\hline
Query & Blue Collar Man (Styx) at 118.1s \\
Answer & Blue Collar Man (REO Speedwagon) at 143.6s \\
Model & MERT \\
Reason & - \\
\hline
\multicolumn{2}{c}{\textbf{Correct Example 2}} \\
\hline
Query & September Gurls (Big Star) at 9.2s \\
Answer & September Gurls (Bangles) at 8.0s \\
Method & Music-domain similarity \\
Reason & Vocal Melody \\
\hline
\multicolumn{2}{c}{\textbf{Incorrect Example 1}} \\
\hline
Query & Gold Dust Woman (Sheryl Crow) at 184.2s \\
Answer & Tomorrow Never Knows (Beatles) at 110.2s \\
Method & Music-domain similarity \\
Reason & Rhythm similarity \\
\hline
\end{tabular}
\caption{Examples of music-level matching segment}
\label{tab:segment_examples}
\end{table}

\section{Conclusion}
In this paper, we suggest a definition of the music plagiarism detection task by three key tasks, distinguishing it from the existing music similarity research. We proposed a comprehensive pipeline based on music segment transcription to perform the music plagiarism detection task. We constructed experiments with the SMP dataset, consisting of actual plagiarism cases with raw audio and their timestamps. This study demonstrates the possibility of a segment-based framework that effectively bridges the gap between theoretical plagiarism detection studies and real-world music applications.

We consider this work to be at an early stage of Music Plagiarism Detection research. The framework built in this study can be extended in various directions. Approaches focused on improving benchmark performance could be developed. New similarity metrics and filtering algorithms specialized in plagiarism detection can be developed by redefining segment-level similarity and music-level similarity. We believe that more advanced models from the MIR research field could be employed with adjustments to suit this specific task.

\bibliographystyle{IEEEbib}
\bibliography{strings,refs}

@article{yuan2023perceptual,
  title={Perceptual and automated estimates of infringement in 40 music copyright cases},
  author={Yuan, Yuchen and Cronin, Charles and Mullensiefen, Daniel and Fujii, Shinya and Savage, Patrick E},
  year={2023},
  publisher={Ubiquity Press}
}

@article{park2022music,
  title={Music plagiarism detection based on siamese cnn},
  author={Park, Kyuwon and Baek, Seungyeon and Jeon, Jueun and Jeong, Young-Sik},
  journal={Hum.-Cent. Comput. Inf. Sci},
  volume={12},
  pages={12--38},
  year={2022}
}

@article{malandrino2022adaptive,
  title={An adaptive meta-heuristic for music plagiarism detection based on text similarity and clustering},
  author={Malandrino, Delfina and De Prisco, Roberto and Ianulardo, Mario and Zaccagnino, Rocco},
  journal={Data Mining and Knowledge Discovery},
  volume={36},
  number={4},
  pages={1301--1334},
  year={2022},
  publisher={Springer}
}

@inproceedings{rouard2023hybrid,
  title={Hybrid transformers for music source separation},
  author={Rouard, Simon and Massa, Francisco and D{\'e}fossez, Alexandre},
  booktitle={ICASSP 2023-2023 IEEE International Conference on Acoustics, Speech and Signal Processing (ICASSP)},
  pages={1--5},
  year={2023},
  organization={IEEE}
}

@inproceedings{kim2023all,
  title={All-in-one metrical and functional structure analysis with neighborhood attentions on demixed audio},
  author={Kim, Taejun and Nam, Juhan},
  booktitle={2023 IEEE Workshop on Applications of Signal Processing to Audio and Acoustics (WASPAA)},
  pages={1--5},
  year={2023},
  organization={IEEE}
}

@inproceedings{donahue2022melody,
  title={Melody transcription via generative pre-training},
  author={Donahue, Chris and Thickstun, John and Liang, Percy},
  booktitle={ISMIR},
  year={2022}
}

@article{chen2019harmony,
  title={Harmony Transformer: Incorporating chord segmentation into harmony recognition},
  author={Chen, Tsung-Ping and Su, Li and others},
  journal={Neural Netw},
  volume={12},
  pages={15},
  year={2019}
}

@article{li2023mert,
  title={Mert: Acoustic music understanding model with large-scale self-supervised training},
  author={Li, Yizhi and Yuan, Ruibin and Zhang, Ge and Ma, Yinghao and Chen, Xingran and Yin, Hanzhi and Xiao, Chenghao and Lin, Chenghua and Ragni, Anton and Benetos, Emmanouil and others},
  journal={arXiv preprint arXiv:2306.00107},
  year={2023}
}

@inproceedings{borkar2021music,
  title={Music plagiarism detection using audio fingerprinting and segment matching},
  author={Borkar, Neetish and Patre, Shubhra and Khalsa, Raunak Singh and Kawale, Rohanshhi and Chakurkar, Priti},
  booktitle={2021 Smart Technologies, Communication and Robotics (STCR)},
  pages={1--4},
  year={2021},
  organization={IEEE}
}

@inproceedings{wang2021preparation,
  title={On the preparation and validation of a large-scale dataset of singing transcription},
  author={Wang, Jun-You and Jang, Jyh-Shing Roger},
  booktitle={ICASSP 2021-2021 IEEE International Conference on Acoustics, Speech and Signal Processing (ICASSP)},
  pages={276--280},
  year={2021},
  organization={IEEE}
}

@inproceedings{marolt2006mid,
  title={A Mid-level Melody-based Representation for Calculating Audio Similarity.},
  author={Marolt, Matija},
  booktitle={ISMIR},
  pages={280--285},
  year={2006},
  organization={Citeseer}
}

@article{serra2009cross,
  title={Cross recurrence quantification for cover song identification},
  author={Serra, Joan and Serra, Xavier and Andrzejak, Ralph G},
  journal={New Journal of Physics},
  volume={11},
  number={9},
  pages={093017},
  year={2009},
  publisher={IOP Publishing}
}

@inproceedings{du2023bytecover3,
  title={Bytecover3: Accurate cover song identification on short queries},
  author={Du, Xingjian and Wang, Zijie and Liang, Xia and Liang, Huidong and Zhu, Bilei and Ma, Zejun},
  booktitle={ICASSP 2023-2023 IEEE International Conference on Acoustics, Speech and Signal Processing (ICASSP)},
  pages={1--5},
  year={2023},
  organization={IEEE}
}

@inproceedings{liu2023coverhunter,
  title={Coverhunter: Cover song identification with refined attention and alignments},
  author={Liu, Feng and Tuo, Deyi and Xu, Yinan and Han, Xintong},
  booktitle={2023 IEEE International Conference on Multimedia and Expo (ICME)},
  pages={1080--1085},
  year={2023},
  organization={IEEE}
}

@article{lu2025melodysim,
  title={MelodySim: Measuring Melody-aware Music Similarity for Plagiarism Detection},
  author={Lu, Tongyu and Geist, Charlotta-Marlena and Melechovsky, Jan and Roy, Abhinaba and Herremans, Dorien},
  journal={arXiv preprint arXiv:2505.20979},
  year={2025}
}

@inproceedings{go2025real,
  title={Real-World Music Plagiarism Detection with Music Segment Transcription System},
  author={Go, Seonghyeon},
  booktitle={2025 Asia Pacific Signal and Information Processing Association Annual Summit and Conference (APSIPA ASC)},
  pages={276--281},
  year={2025},
  organization={IEEE}
}

@inproceedings{chang2021neural,
  title={Neural audio fingerprint for high-specific audio retrieval based on contrastive learning},
  author={Chang, Sungkyun and Lee, Donmoon and Park, Jeongsoo and Lim, Hyungui and Lee, Kyogu and Ko, Karam and Han, Yoonchang},
  booktitle={ICASSP 2021-2021 IEEE International Conference on Acoustics, Speech and Signal Processing (ICASSP)},
  pages={3025--3029},
  year={2021},
  organization={IEEE}
}

@article{gaillard2017large,
  title={Large scale reverse image search},
  author={Gaillard, Mathieu and Egyed-Zsigmond, El{\H{o}}d},
  journal={XXXV{\`e}me Congr{\`e}s INFORSID},
  volume={127},
  year={2017}
}

@article{ellis2007covers80,
  title={The" covers80" cover song data set},
  author={Ellis, Daniel PW},
  journal={URL: http://labrosa. ee. columbia. edu/projects/coversongs/covers80},
  year={2007}
}

@inproceedings{liu2023fine,
  title={Fine-grained music plagiarism detection: Revealing plagiarists through bipartite graph matching and a comprehensive large-scale dataset},
  author={Liu, Wenxuan and He, Tianyao and Gong, Chen and Zhang, Ning and Yang, Hua and Yan, Junchi},
  booktitle={Proceedings of the 31st ACM International Conference on Multimedia},
  pages={8839--8848},
  year={2023}
}

\end{document}